\documentclass[conference]{IEEEtran}
\IEEEoverridecommandlockouts
\usepackage{cite}
\usepackage{amsmath,amssymb,amsfonts}
\usepackage{algorithmic}
\usepackage{graphicx}
\usepackage{subcaption}
\PassOptionsToPackage{hyphens}{url}
\usepackage{hyperref}
\usepackage[table,xcdraw]{xcolor}
\usepackage{textcomp}
\def\BibTeX{{\rm B\kern-.05em{\sc i\kern-.025em b}\kern-.08em
    T\kern-.1667em\lower.7ex\hbox{E}\kern-.125emX}}
\begin{document}

\title{Processing Particle Data Flows with SmartNICs}

\author{\IEEEauthorblockN{Jianshen Liu}
\IEEEauthorblockA{\textit{Comput. Sci. \& Engineering} \\
\textit{UC Santa Cruz}\\
Santa Cruz, CA, USA \\
jliu120@ucsc.edu}
\and
\IEEEauthorblockN{Carlos Maltzahn}
\IEEEauthorblockA{\textit{Comput. Sci. \& Engineering} \\
\textit{UC Santa Cruz}\\
Santa Cruz, CA, USA \\
0000-0001-8305-0748}
\and
\IEEEauthorblockN{Matthew L. Curry}
\IEEEauthorblockA{\textit{Scalable System Software} \\
\textit{Sandia National Laboratories}\\
Albuquerque, NM, USA \\
mlcurry@sandia.gov}
\and
\IEEEauthorblockN{Craig Ulmer}
\IEEEauthorblockA{\textit{Scalable Modeling \& Analysis} \\
\textit{Sandia National Laboratories}\\
Livermore, CA, USA \\
cdulmer@sandia.gov}
}

\maketitle

\begin{abstract}
Many distributed applications implement complex data flows and need a flexible mechanism for routing data between producers and consumers. Recent advances in programmable network interface cards, or SmartNICs, represent an opportunity to offload data-flow tasks into the network fabric, thereby freeing the hosts to perform other work. System architects in this space face multiple questions about the best way to leverage SmartNICs as processing elements in data flows. In this paper, we advocate the use of Apache Arrow as a foundation for implementing data-flow tasks on SmartNICs. We report on our experiences adapting a partitioning algorithm for particle data to Apache Arrow and measure the on-card processing performance for the BlueField-2 SmartNIC. Our experiments confirm that the BlueField-2's (de)compression hardware can have a significant impact on in-transit workflows where data must be unpacked, processed, and repacked.
\end{abstract}


\section{Introduction}
Distributed applications routinely need an efficient mechanism for routing data from collections of producers to collections of consumers. For example, in high-performance  computing (HPC), scientific simulation workflows generate parallel streams of output that data management services gather, analyze, and transform before archiving to storage. Similarly, geographic information system (GIS) data flows often aggregate sensor data from a variety of sources, reorganize it, and then transmit derived data products to remote subscribers based on user specifications. Current-generation systems implement their distributed data-flow operations in host-level software that communicates through standard network protocols.

Commercial hardware vendors have recently started supplementing their products with embedded processing resources that may offer significant advantages for systems that implement distributed data flows. Storage vendors offer computational storage devices (CSDs) with user-programmable FPGAs and processors that allow the user to execute computations at the disk. These resources allow filtering operations to be placed close to data sources and may dramatically reduce the amount of data returned to the user. In the networking arena, vendors are placing embedded processors in network interface cards (NICs). These \textit{SmartNICs} can help process in-transit data and are especially useful in distribution tasks where data flows must transmit customized data products to multiple destinations.

The presence of these programmable, embedded processors throughout the system architecture motivates the use of eusocial methods, where many low-capability devices are programmed to collectively act towards a more complex, system-level goal~\cite{kufeldt2018eusocial}. Our work in eusocial methods is focused on constructing general-purpose data management software that makes it easier for system developers to map different data flows onto a collection of embedded processors in the network and storage fabrics. There are many questions architects face in this space. How should in-transit data be represented and processed? How well do current embedded processors perform fundamental data-sifting operations? When is data compression profitable in these architectures?

In this paper, we focus on constructing a computational engine that can perform a variety of eusocial tasks on SmartNICs. We advocate for the use of Apache Arrow~\cite{apache_arrow} to standardize how in-transit data is represented and processed by eusocial devices. As a means of better evaluating current generation hardware, we have implemented a partitioning algorithm for particle datasets on the NVIDIA BlueField-2 VPI SmartNIC~\cite{liu2021performance} and measured its performance with different datasets. Additional experiments with hardware accelerators on this card confirm that compression hardware can have a significant impact on performance and motivate the need for Arrow enhancements in future work.

\section{Particle Data Flows}

Many HPC and GIS applications operate on particle data and rely on complex data management services to route particle state information between producers and consumers distributed throughout the network. While particle datasets are smaller in size than multimedia datasets, application data flows can be challenging to implement because the datasets contain many small items that are tedious to inspect. As such, it is beneficial to consider hardware environments that can offload the task of reorganizing and sifting through the data. In this section we provide application examples from the HPC and GIS spaces, and discuss a common data-flow use case where data is transitioned from a spatially-organized form to a temporally-organized form.

\subsection{Particle Simulations in Scientific Computing}

Many simulations in scientific computing employ particle-in-cell (PIC) methods~\cite{harlow1964particle, arber2015contemporary, derouillat2018smilei} to model different phenomena. These simulations manage a large collection of discrete particles and track their progress as they move through time and space. A particle is defined by a small amount of state information, such as its position, velocity, charge, and type. Given that simulation fidelity improves as the number of particles in the simulation increases, researchers typically leverage parallel simulation techniques to distribute the data and work across many compute nodes. Although a simulator's particle data may contain a treasure trove of information for scientists, the sheer size of this data makes it infeasible to save except in the case of occasional checkpointing. Analysts may run supplemental analysis applications in parallel with the simulation and inspect subsets of the data without impeding the simulation.

\subsection{Asset Tracking with Geographic Information Systems}

A similar need to collect and process large amounts of particle data can be found in GIS applications that manage sensor data about real-world assets, such as airplanes, ships, and land vehicles. Although the ``particles'' in these systems are significantly larger in physical size than those in the simulations, the processing challenges for manipulating the data flows are the same: different distributed sensor systems produce continuous feeds of observation data that needs to be reorganized to be of value to downstream consumers.

\subsection{Reorganizing Spatial Indices to Temporal Indices}

\begin{figure}[ht]
  \includegraphics[width=\columnwidth]{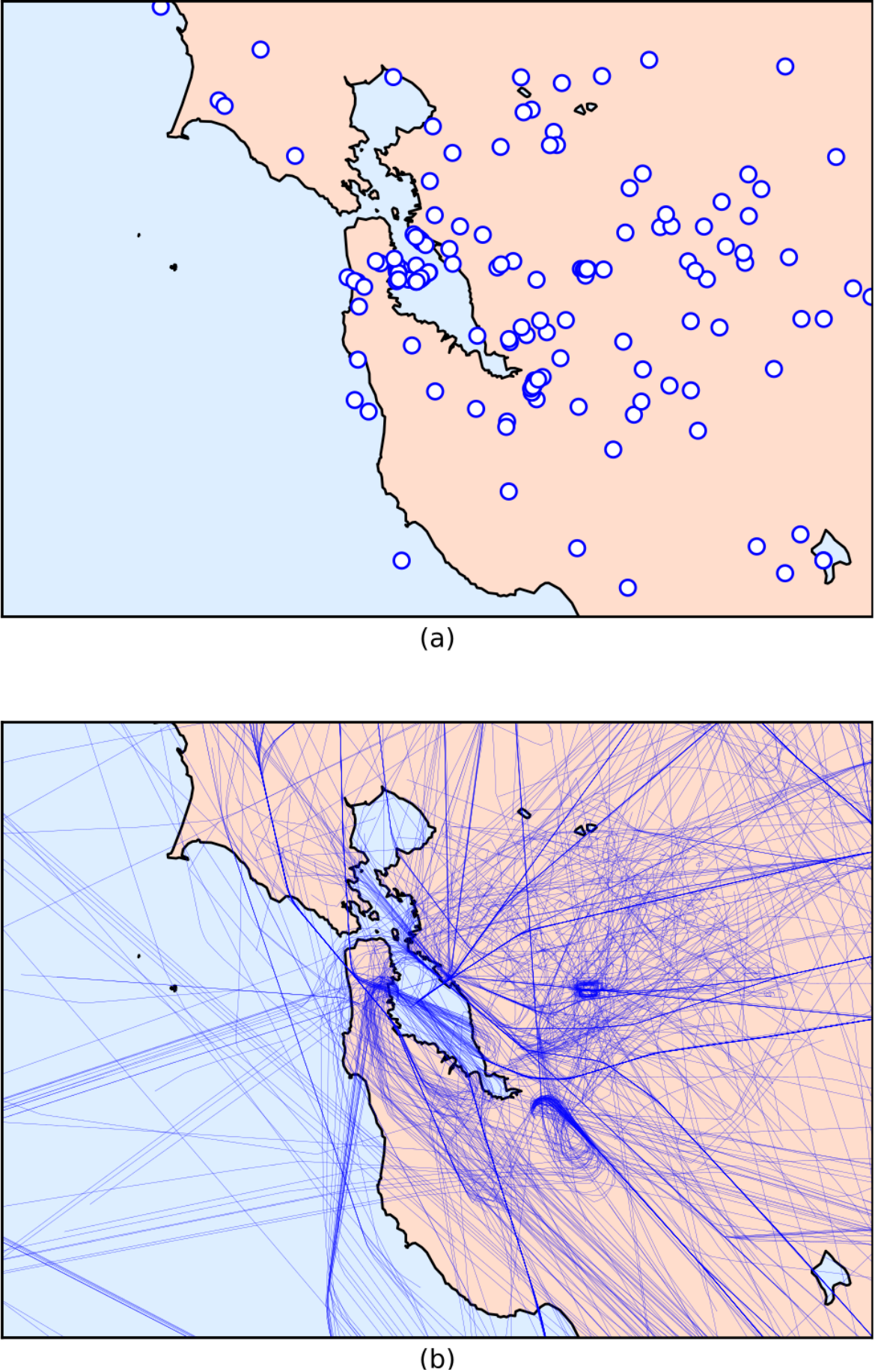}
  \caption{Airplane position data in a (a) point form for a snapshot in time and (b) track form for a window of one hour}
  \label{fig:airplane_tracks}
\end{figure}

One hardship of working with particle data flows is that there are significant differences between the way producers and consumers expect data to be organized. Producers typically organize data in a spatial manner, where each sensor generates an update for all items in a physical region during a particular time interval. In contrast, analytics consumers often need data organized temporally for each item. As illustrated in Fig.~\ref{fig:airplane_tracks} with airplane data, temporal tracks (b) can yield better insight into patterns of activity than positional snapshots (a) alone.

Distributed, log-structured merge (LSM) trees~\cite{picoli2019lsm} are a convenient mechanism for converting particle data flows from a spatial organization to a temporal organization when multiple producers and consumers are involved. In this approach, a collection of processing elements is used to reorganize data as it moves through different stages in the tree. Each processing element absorbs incoming data until it reaches its storage capacity. When compaction is necessary, data is split based on particle IDs and transmitted to the next appropriate processing element in the tree. While processing elements only need to store, sift, and transmit blocks of particles, they can greatly improve the searchability of online datasets.

\section{Smart Network Interface Cards (SmartNICs)}

Over the last decade, hardware vendors have introduced programmable network interface cards or SmartNICs that enable users to place custom computations at the edge of the network fabric. While the original motivation for developing SmartNICs was to allow security researchers to monitor and inspect network flows in real time~\cite{le2017uno, sabin2015security}, the need for tighter infrastructure control in cloud computing platforms has driven SmartNIC vendors to create more powerful cards. Vendors such as NVIDIA (formerly Mellanox), Fungible, Chelsio, Intel, and Xilinx have constructed SmartNICs that allow users to embed computations at the network's edge. While some SmartNIC architectures employ FPGAs or ASICs to maximize packet processing performance, most feature a multicore embedded processor that is easier for developers to leverage.

\subsection{NVIDIA BlueField-2 VPI SmartNIC}

The NVIDIA BlueField-2 VPI SmartNIC is a programable NIC that features eight A72 ARM processor cores, 16GB of DRAM, and 60GB of flash storage. It includes two 100Gb/s network ports that can interact with either InfiniBand or Ethernet. The default software stack for the card boots an Ubuntu 20.04 installation of Linux from on-card eMMC storage. The ARM processors can be configured to either (1) intercept traffic between the host and network or (2) serve as a separate host that shares access to the network ports. The BlueField-2 includes three hardware acceleration units to improve application performance: a RegEx engine for data filtering, a SHA-2 unit for encryption functions, and a (de)compression engine.

\subsection{BlueField-2 Compression Accelerator}

Data compression is important in data-intensive applications because it reduces the amount of data that needs to be transmitted through the network, cached in memory, and stored on disk. Most big data I/O libraries (e.g., Avro~\cite{vohra2016apache_avro}, Parquet, ORC~\cite{vaddeman2016data}, and Arrow IPC) feature built-in compression support for a variety of codecs. As such, any application that processes this data must be capable of decompressing and compressing the data in a manner that is compliant with the library's data format.

The BlueField-2's compression accelerator can efficiently compress and decompress data using the DEFLATE algorithm~\cite{deutsch1996deflate}. DEFLATE is widely used and is a key part of standards such as PNG~\cite{boutell1997png}, HTTP~\cite{mogul2002delta}, TLS~\cite{hollenbeck2004transport}, and SSH~\cite{ylonen2006secure}. The BlueField-2's hardware implementation of DEFLATE is compatible with the zlib library, which means that when needed, data compressed by the hardware can be decompressed by software and vice versa. This interoperability is critical in eusocial applications because it enables data processing tasks to be ``pushed down'' to offload the host or ``pushed back'' in situations where an embedded device becomes saturated with work. The BlueField-2's compression hardware is currently accessed through the Data Plane Development Kit (DPDK)~\cite{cerrato2014supporting,dpdk}, which is a library for constructing high-performance data-plane applications on top of a variety of network hardware devices. The compression hardware is designed to process a stream of individual data packets in an efficient manner and includes DMA hardware to facilitate the movement of data between the accelerator and memory. Compression performance for the BlueField-2 hardware is discussed in Section~\ref{sec:compression_experiments}.

\subsection{Performance Expectations for SmartNICs}

It is important to recognize that the processing resources available to a SmartNIC are considerably less than those available to a host. In prior benchmark experiments~\cite{liu2021performance}, we observed that the BlueField-2's ARM processors performed roughly an order of magnitude slower than host systems, due to a combination of processor and memory bandwidth limitations. The gap between embedded and server processors is \textit{not unexpected} and is unlikely to change in the foreseeable future.

While SmartNICs are not general-purpose accelerators, there are several scenarios where we expect the hardware to be beneficial to applications. First, SmartNICs are sufficient for performing simple filtering and projection operations that do not involve complex computations. Second, data flows involving compression or encryption may be able to leverage the card's hardware accelerators and achieve a speedup. Finally, it may be advantageous to offload low-rate, asynchronous event processing to a SmartNIC, due to the disturbances these operations have on other tasks that run on the host.

\section{Software Infrastructure For In-transit Processing}

Data-intensive applications in both science and commercial enterprises are often constructed using multiple systems with independent implementation histories and choices of programming languages. A key operating expense of these applications is the movement of data across these systems. But what sounds like a problem of moving data between systems is really the challenge of efficiently (1) converting the data from a system's internal in-memory representation to a wire format and (2) accessing large amounts of data via record-by-record API calls. This is precisely the challenge that Apache Arrow set out to address, an open-source project that since 2016 has been quickly gaining adoption within the data science community.

\subsection{Apache Arrow}

The key insight underlying the design of Apache Arrow ecosystem is that by creating an efficient open data processing platform around a common and efficient in-memory data representation with many different programming language bindings (so far C, C++, C\#, Go, Java, JavaScript, Julia, MATLAB, Python, R, Ruby, and Rust), data can move efficiently between the ecosystem's data processing engines running on different systems. Data processing and exchange can be implemented with a number of building blocks that include the Parquet file format~\cite{vohra2016apache_parquet}, the Flight framework for efficient data interchange between processes~\cite{mckinney2019introducing}, the Gandiva LLVM-based JIT computation for executing analytical expressions by leveraging modern CPU SIMD instructions to process Arrow data~\cite{pindikura_introducing_2018}, and Awkward Array for restructuring computation on columnar and nested data~\cite{pivarski2020awkward}. On top of these building blocks exist a number of Arrow integration frameworks, including the Fletcher framework that integrates FPGAs with Apache Arrow~\cite{peltenburg2019fletcher}, NVIDIA's RAPIDS cuDF framework that does similar for GPUs~\cite{raschka2020machine,rapids}, the Plasma high-performance shared-memory object store~\cite{nishihara_plasma_2017}, the Skyhook distributed storage plug-in to embed Arrow processing engines within Ceph storage objects~\cite{chakraborty2022skyhook, chakraborty_skyhook_2022}, and the Substrait effort to standardize an open format for query plans between query optimizers and processing engines~\cite{nadeau_substrait}. There are many more projects that are adopting the Apache Arrow in-memory representation and the Dataset Interface that abstracts over a variety of file formats and other data sources~\cite{apache_arrow_dataset_interface}. The amount of significant investment poured into this ecosystem is reflected by its recent cadence of four major version releases per year, most recently version 9.0.0 with 1061 issues resolved by 114 distinct contributors over 3 months.

\subsection{Data Organization in Apache Arrow and Opportunities}

Apache Arrow represents tabular data in a columnar, randomly-accessible in-memory format, allowing for nested data structures and null values. The format is designed to maximize CPU throughput by optimizing the data layout for pipelining, SIMD instructions, and cache locality. Data is communicated by schema information involving one or more optional metadata dictionary batches that are followed by record batches. A record batch is composed of multiple arrays, each representing a part of the data from one or more fields from a table. Record batches are designed to be the unit of data processing communicated to and from processing engines. Batching of records minimizes the need for record-based API calls and the batch size can be optimized for pipeline processing while the columnar layout allows for SIMD instructions~\cite{nadeau_vectorized_2019}.

\begin{figure}[ht]
  \includegraphics[width=\columnwidth]{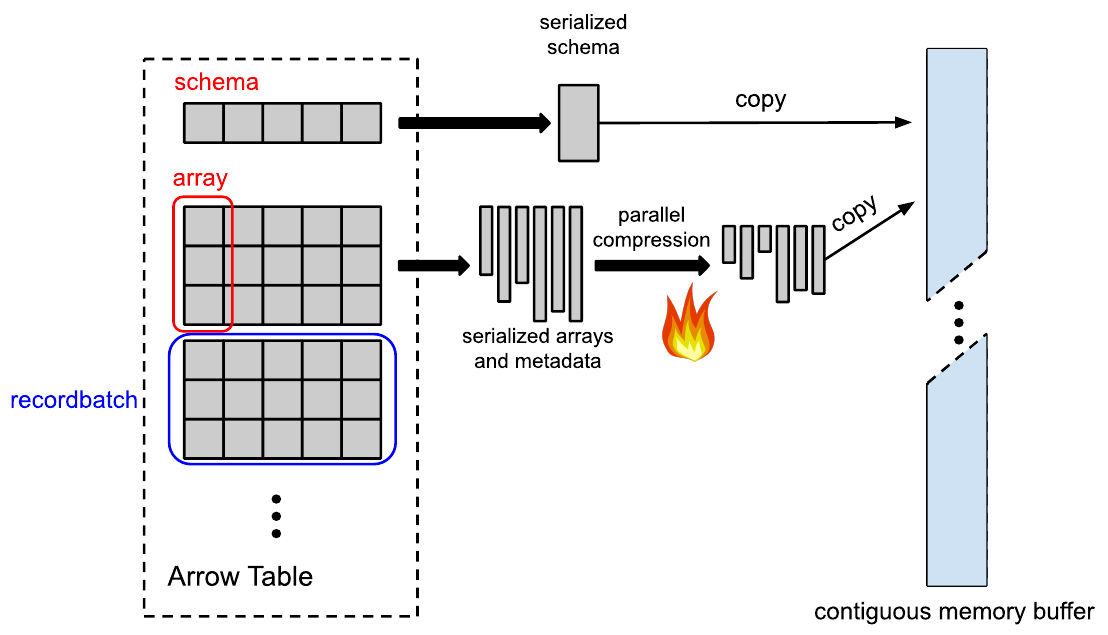}
  \caption{A simplified data serialization process of an Arrow table}
  \label{fig:serialize_arrow_tables}
\end{figure}

Arrow IPC format is a protocol that encodes record batches into contiguous bytes for storing in either files or memory. This encoding process is known as serialization. Fig.~\ref{fig:serialize_arrow_tables} shows how a typical Arrow table is serialized into a byte sequence in the IPC format.

The schema of a table is first serialized and written to the output memory buffer. Then, for each record batch, the arrays it contains will be serialized one after another according to their types. With all record batches being serialized resulting in a buffer vector, these buffers will be compressed in parallel for better compression throughput. The number of threads that will be spawned in this process typically matches the size of the buffer vector. For most array types, serializing an array produces two buffers --- a data buffer and an additional buffer containing the metadata called the validity bitmap. As a result, the total number of threads started is a multiple of the number of columns, which can keep many cores busy. For example, in one of our datasets for experiments, the loaded Arrow table contains 17 columns; however, the compression phase spawns 35 threads occupying more cores than the ones available from a CPU socket. As such, the generated compression workload may hinder or stall performance-critical applications such as simulations that are running on the same host. Leveraging the compression accelerator from the BlueField-2 SmartNIC provides an opportunity to break the dependence of compression performance on intensive computing resource occupation.

\subsection{Related Work}

While Apache Arrow meets several of our needs, our work can be adapted to other important data management libraries. HDF5~\cite{gu2019hdf5} is an established library for representing scientific data in stored data. Although it does not include a rich set of primitives for dispatching queries on in-memory data, it does provide a modular interface for extending the library's capabilities. Kokkos~\cite{edwards2013kokkos} is a computational library that aims to provide performance portability across different data-parallel architectures. Its data views provide a simple structure for hosing data vectors in a way that simplifies transport. Similarly, VTK-m~\cite{moreland2016vtk} is a library for facilitating data-parallel visualization operations.

\section{Data Partitioning Experiments}

As a means of better evaluating the suitability of Apache Arrow for processing particle data flows on SmartNICs, we implemented a data partitioning algorithm used in LSM trees. In this work, we used Apache Arrow to represent particles in a tabular form that is suitable for transfer over the network and leveraged Arrow's filtering operations to split a table into smaller tables based on particle IDs. We measured the amount of time required to unpack, partition, and repack data for three particle datasets from different communities to demonstrate the flexibility of this approach.

\subsection{Implementation}

We constructed a C++ program that inspects and processes in-transit data objects in network data flows. This program is supplied with a contiguous-memory data object and is expected to provide one or more contiguous result objects that are to be sent to different locations. For this work we use Apache Arrow's IPC methods to handle transformations between a serialized object that can be transported in the network and an in-memory format that is suitable for tabular computations.

The partitioning algorithm examines a table and uses a small number of bits in the particle ID field to determine which output table should hold each particle. Although Arrow provides a group-by function that would be useful for performing a split in a single pass, it is currently limited to statistical operations. As such, we implemented the partitioning as a multistep algorithm that executes a select query to generate each table. While far from ideal, this approach is acceptable in the LSM tree work because of the low-fanout requirements of the distributed algorithm.

\subsection{Reference Datasets}

Three particle datasets were used in these experiments to provide better insight into the performance of the algorithm with different data:
\begin{itemize}
\item \textbf{TrackML Particle Tracking Challenge} (``Particles'')~\cite{amrouche2020tracking}: CERN supplied a particle simulation dataset for a machine learning competition hosted through Kaggle in 2018. This dataset contains 10 numerical fields per particle.
\item \textbf{OpenSky Network} (``OpenSky Planes'')~\cite{schafer2014bringing}: The OpenSky Network collects worldwide ADSB information for airplanes from volunteers. Entries contain 16 fields composed of a mix of numerical and string values.
\item \textbf{NOAA Maritime} (``Ships'')~\cite{national_oceanic_and_atmospheric_administration_vessel}: NOAA provides historical AIS position data for ships near the US coastline. Daily data was converted to a particle format that contained 17 fields composed of a mix of numerical and string values.
\end{itemize}

Given that the BlueField-2 SmartNIC operates with 16GB of DRAM, we set a 1GB limit for the size of uncompressed data to use in our experiments. We decompressed each dataset, selected the number of rows that would be closest to 1GB in size, and then recompressed the data to serve as input to the experiments.

\subsection{Experiments}

\begin{figure}[ht]
  \includegraphics[width=\columnwidth]{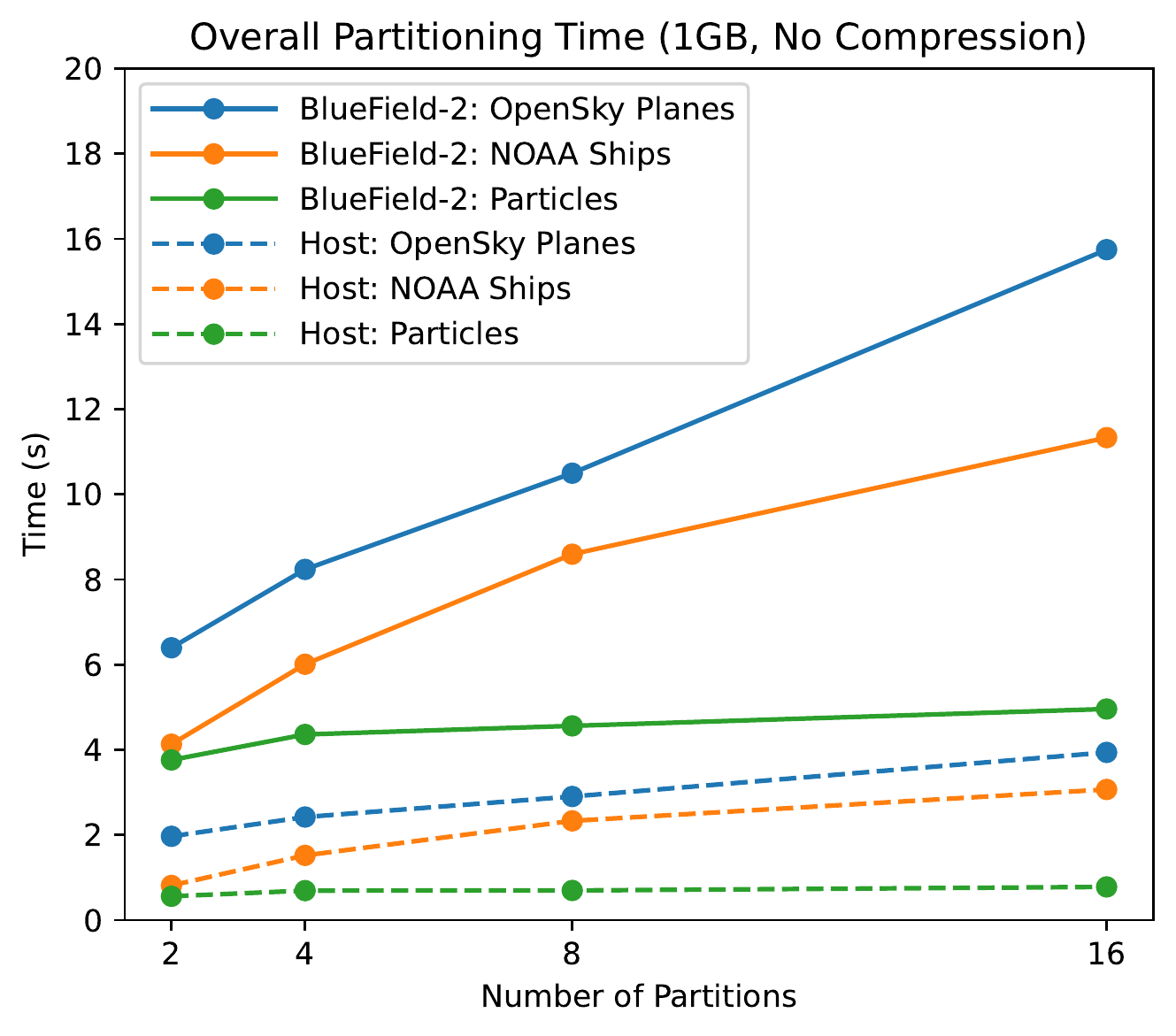}
  \caption{Overhead for partitioning without compression}
  \label{fig:overhead_for_partitioning}
\end{figure}

Performance experiments were conducted on a compute node that features a 32-core AMD EPYC 7543P processor and a BlueField-2 VPI card. In the first experiment, we measured the overall amount of time required for the host or SmartNIC to unpack, partition, and repack the tabular data into 2 to 16 output partitions. As depicted in Fig.~\ref{fig:overhead_for_partitioning}, the host operates roughly four times faster than the BlueField-2 when processing uncompressed data. Increasing the number of partitions increased the processing time in most cases. A closer inspection of the ``Particles'' dataset revealed an ID address space issue that resulted in a distribution imbalance. These issues can be mitigated by hashing the ID or selecting ranges that are more meaningful to the application.

\begin{figure}[ht]
  \includegraphics[width=\columnwidth]{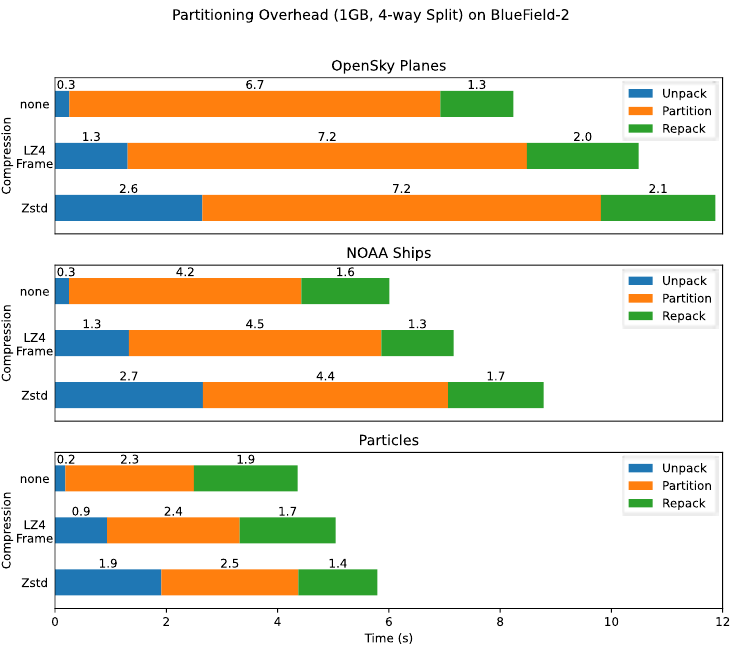}
  \caption{Timing breakdown for a 4-way split on the BlueField-2 using different software compression methods}
  \label{fig:timing_breakdown_for_split}
\end{figure}

The second experiment examines the impact of Apache Arrow's built-in software compression mechanisms on performance. These tests vary whether the input and output objects are serialized with no compression, LZ4 Frame compression~\cite{collet_lz4_2020}, or Zstd compression~\cite{collet2018zstandard}. Fig.~\ref{fig:timing_breakdown_for_split} provides the timing breakdowns for unpacking, partitioning, and repacking 1GB of particle data when performing a 4-way split. As expected, uncompressed data is significantly faster to read than compressed data. Repacking the data, however, is similar in all cases. This overhead highlights the fact that serialization by itself is an expensive operation.

\begin{figure}[ht]
  \includegraphics[width=\columnwidth]{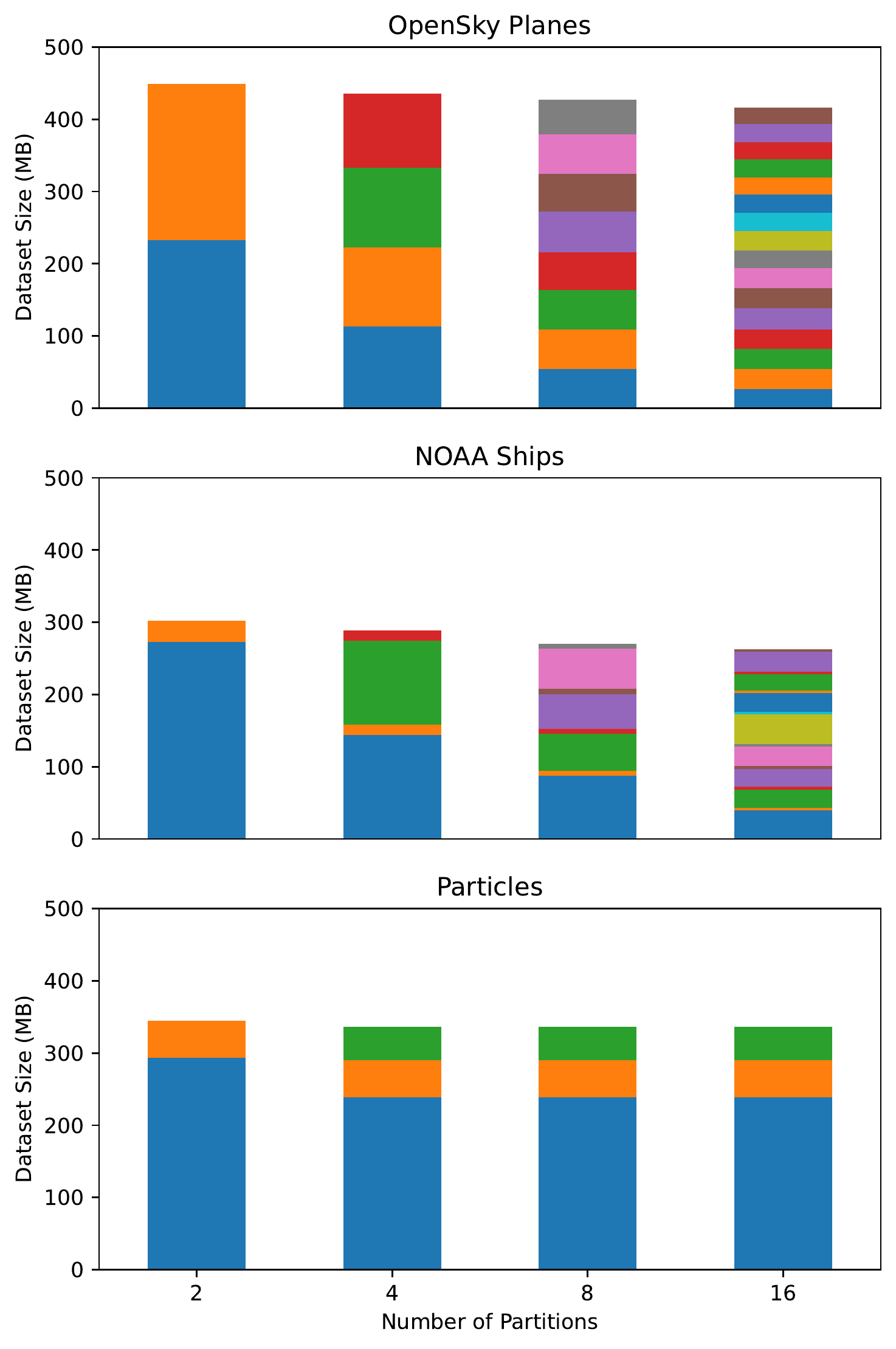}
  \caption{Aggregate dataset sizes when varying the number of partitions and compressing with Zstd}
  \label{fig:compression_splits}
\end{figure}

Examining the output sizes of the individual, serialized partitions generated in the second experiment provides greater insight into how partitioning affects compression results. Fig.~\ref{fig:compression_splits} provides a breakdown of how large each output partition is when using Zstd compression and the lowest 1 to 4 bits of the particle ID to split the three input datasets. In the OpenSky Planes dataset, the lower bits of the ID are diverse and yield equally-sized ouput partitions. There is a slight decrease in the aggregate size of the output data as the number of partitions increases because the individual partitions have more data redundancy that the compression algorithm can exploit.

In contrast, the NOAA Ships and the Particles datasets have less diversity in the lower bits of the particle ID field. As such, the partitioning algorithm splits the data into uneven portions. This property is undesirable because it may create load balancing issues with downstream consumers of this data. While the aggregate size of the NOAA Ships dataset improves as the number of partitions increases, the Particles dataset does not as its IDs can only be split into three partitions. These examples indicate that it is worthwhile for architects to understand the characteristics of their data and select partition address bits that will result in balanced outputs.

\subsection{Discussion}

While the host processors in this experiment yielded better performance, it is important to note that the BlueField-2's embedded processors were performant enough to be of value in many data flows. Scenarios where producers generate periodic bursts of data are applicable, as the SmartNIC can absorb the bursts and process the data before the next wave arrives.

Implementing the partitioning operation with Apache Arrow highlighted its development advantages. Arrow's well-reasoned data primitives and existing support for serialization, compression, and processing greatly simplified the implementation effort. Our implementation worked with all three datasets without modification, even though each dataset had different data components and ID bitwidths. Although the current version of Arrow does not have all the primitives of a higher-level library such as Pandas~\cite{mckinney2011pandas}, it contains adequate primitives to implement a variety of operations.

\section{Compression Experiments}\label{sec:compression_experiments}

As demonstrated in the previous section, converting between on-the-wire and in-memory formats is an important and time-consuming task for systems that process in-transit tabular data. Given that the Bluefield-2 SmartNIC provides a compression accelerator and multiple cores that Apache Arrow can leverage, it is worthwhile to explore the different compression options that are available for packing and unpacking data. We conducted three experiments to answer each of the following questions: (1) Is the compute overhead caused by software-based compression significant enough to justify offloading the (de)compression to hardware accelerators? (2) How does the throughput performance of hardware-based compression compare with software-based compression in a threaded environment? (3) Is there a change in the compression ratio between the hardware- and software-based methods?

\subsection{Compression Hardware Challenges}

The BlueField-2's compression hardware can be accessed through the Data Plane Development Kit (DPDK) library. Unfortunately, this library is highly tuned for network operations and is organized around a packet-processing model that can be cumbersome for other types of applications. We faced several challenges in adapting DPDK's compression functions to process our Arrow data. First, individual data packets have a maximum size of 64KB. To compress larger amounts of data, developers must slice input and output buffers into packet-sized segments and then generate a packet that contains a list of compression commands for processing each segment. Second, converting between contiguous and segmented data representations can result in extra memory allocations and copies that disrupt the throughput of the data flow through the compression hardware. Optimizing the pipeline requires a detailed understanding of both DPDK and the hardware, and is tedious for users that simply want to (de)compress large blocks of data. Third, embedded hardware environments have limited resources. Therefore, recycling resources after each compression operation (while still managing errors) is extremely important. Finally, a single ARM CPU core may not be sufficient for maximizing the performance of the compression accelerator. As such, it is valuable to construct a pipeline that pre-allocates memory and divides work among cores as needed.

\subsection{Implementation: Bitar}

To simplify accessing the compression hardware for data compression, we implemented the \emph{Bitar}~\cite{liu_simplify_accessing_hardware_2022} library on top of DPDK and Arrow. Bitar provides a convenient (de)compression API and features zero-copy processing, synchronous and asynchronous operation, and multicore/multidevice support. It is specifically designed to operate without root privileges, which is uncommon in DPDK-based applications. Bitar also allows users to access the BlueField-2's compression hardware from either the host's or BlueField-2's processors.

\subsection{Experiments}

All experiments in this section were carried out on a CloudLab~\cite{duplyakin2019design} host that has two AMD EPYC 7542 CPUs (a total of 64 cores), 512GB of DDR4 memory, and a BlueField-2 SmartNIC connected with PCIe 4.0 x16 lanes. Each experiment was run on all three reference datasets with a maximum outstanding data window size of 160MB due to memory constraints imposed by DPDK and the pipelined nature of the compression hardware.

Since Bitar has not yet been fully integrated into Arrow, our experiments compress Arrow tables differently depending on whether software- or hardware-based compression is measured. The software-based approach relies on Arrow's existing compression mechanisms, which serialize and compress each column independently before writing the final output buffer (i.e., ``inner compression''). In contrast, the hardware-based approach serializes the entire table and then streams the data through the compression hardware (i.e., ``outer compression''). While the former is preferred, the latter is sufficient for network transfers. Furthermore, comparing the performance of these approaches can help determine the benefits of integrating hardware compression into Arrow.

\subsubsection{Software Compression Overhead for a Single Thread}

Our first research question focuses on whether software-based compression overhead is significant enough to justify hardware acceleration. To answer this question, we constructed an experiment that measures the amount of time for a single thread to pack and unpack Arrow data in software using different codecs.  We intentionally excluded the memory allocation time in this experiment given that it can be preallocated using historical knowledge of output buffer sizes.

\begin{figure}[ht]
  \includegraphics[width=\columnwidth]{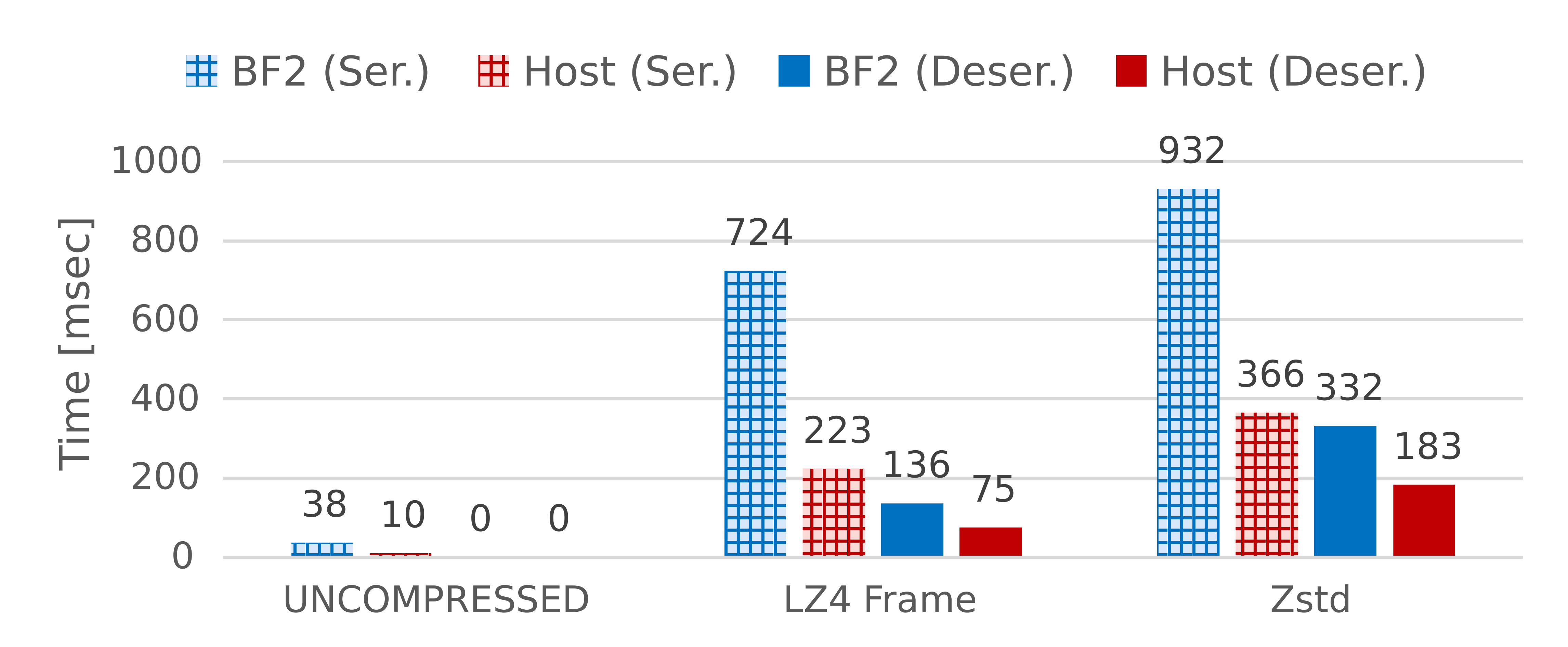}
  \caption{Single-thread (de)serialization time with different compression codecs}
  \label{fig:single-thread_(de)serialization_time}
\end{figure}

Timing results for the ``Particles'' dataset (see Fig.~\ref{fig:single-thread_(de)serialization_time}) indicate that (de)serialization without compression is efficient, thanks to the zero-copy buffer design of Arrow's IPC format. However, involving either LZ4 Frame or Zstd compression introduces significant CPU overhead and increases time consumption by one to two orders of magnitude. For example, serialization without compression on the host takes 10 milliseconds, while adding LZ4 Frame compression to the serialization increases the time to 223 milliseconds. We observed similar results using the other two reference datasets. Given that compression is a significant impediment to performance, we conclude that acceleration is worthwhile in performance-sensitive applications.

\subsubsection{Throughput in a Threaded Environment}

Our second question focuses on how well the software- and hardware-based compression methods perform in a threaded environment. One advantage of Arrow is that it automatically parallelizes the packing and unpacking of tables by dispatching each column's work to its own thread. In Bitar's case, multiple threads can be used to maximize the amount of work supplied to the compression hardware. Since the (de)compression is part of the (de)serialization process in Arrow, we conducted experiments to observe how the (de)serialization throughput improves when scaling (de)compression to use an optimal number of worker threads.

\begin{figure}[ht]
  \includegraphics[width=\columnwidth]{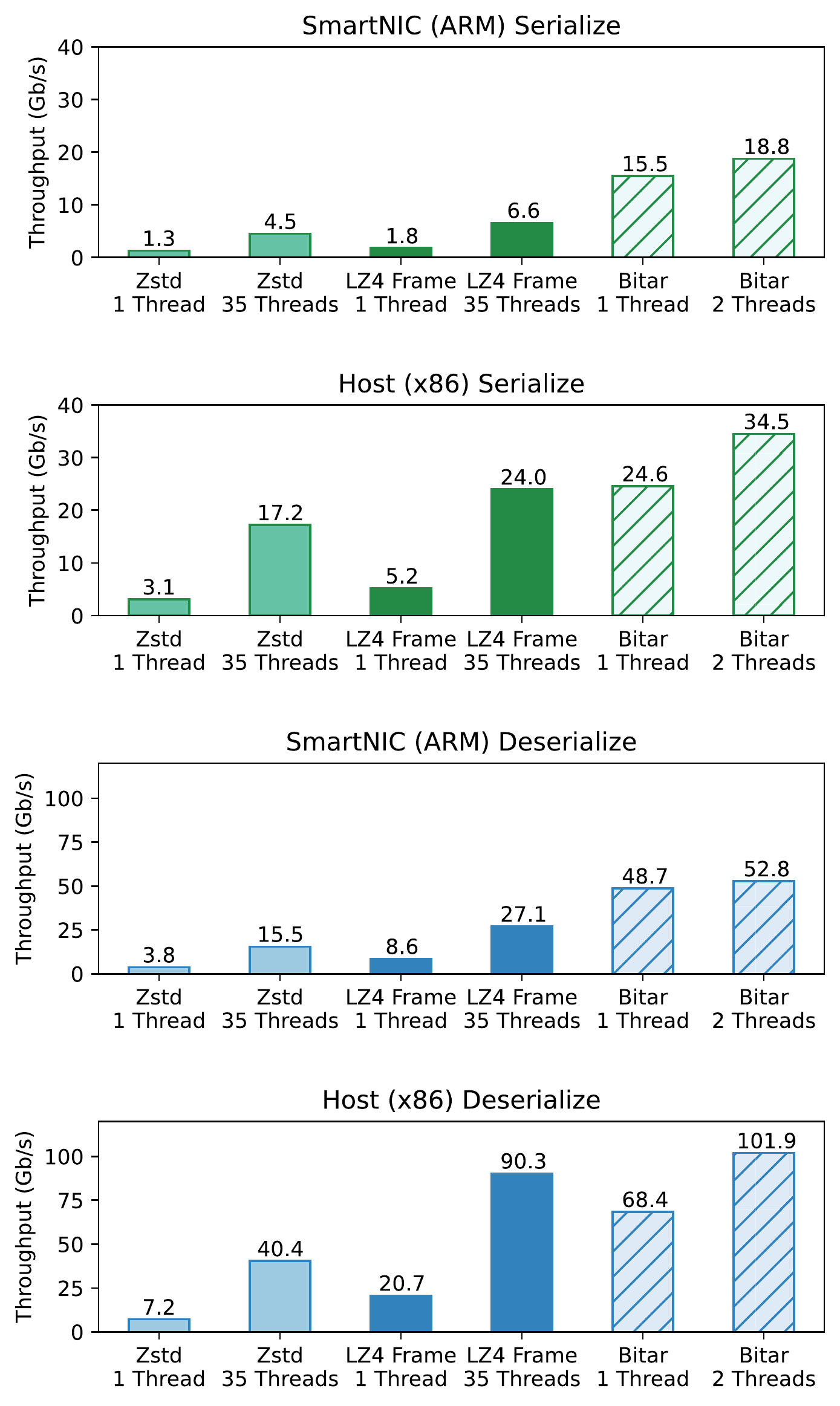}
  \caption{(De)serialization throughput with different compression codecs and degrees of parallelism}
  \label{fig:bitar-paper}
\end{figure}

\begin{figure}[ht]
  \includegraphics[width=\columnwidth]{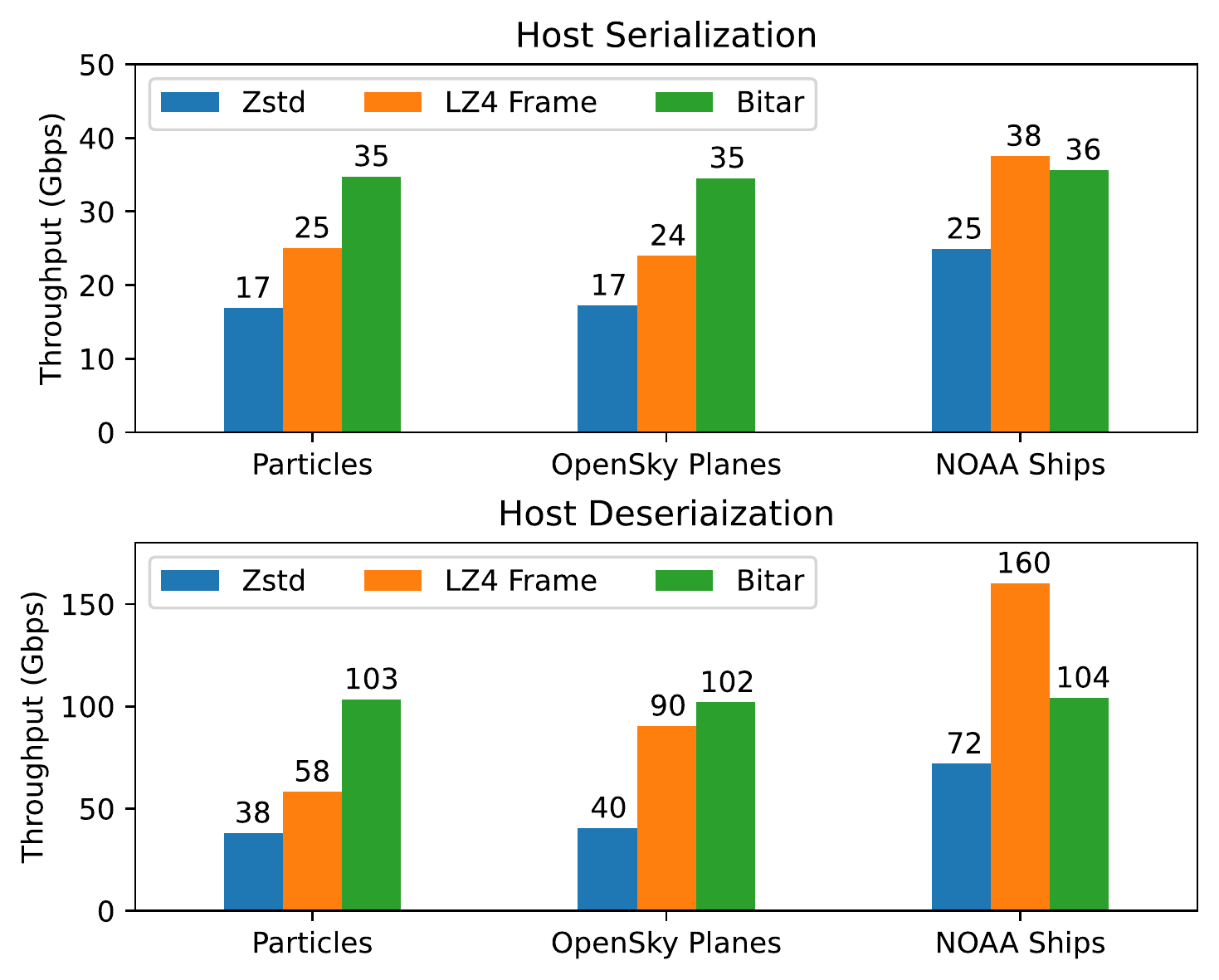}
  \caption{The maximum throughput performance on the host for all three datasets}
  \label{fig:bitar-host-max}
\end{figure}

Fig.~\ref{fig:bitar-paper} shows the throughput measurements for the ``OpenSky Planes'' reference dataset. Without limiting the number of threads in the experiment, both LZ4 Frame and Zstd used 35 threads during compression and decompression. In contrast, the hardware compression throughput with Bitar was maximized when using only two threads, as we did not see higher throughput with more threads. Note that, due to the slower memory subsystem of the SmartNIC, the serialization throughput with Bitar on the host is higher than that on the SmartNIC. In general, for this dataset Bitar outperformed software-based compressions in all cases. The maximum throughput on the host with different codecs for each of the three datasets is summarized separately in Fig.~\ref{fig:bitar-host-max}. To better illustrate the advantages of using the hardware accelerator for (de)compression, we list the (de)serialization speedup with Bitar in Table~\ref{tab:serialization_speedup} and~\ref{tab:deserialization_speedup}. For compression with a single thread on the host, depending on the codec and dataset used, serialization with Bitar can achieve between 4.6-8.6x higher throughput than serialization with software-based compressions. For compression with multiple threads, the use of Bitar can speed up the serialization throughput on the host by 1-2x. For decompression with a single thread on the host, using Bitar can speed up throughput by 3.3-10.8x. For multithreaded decompression, Bitar outperformed ZSTD in all cases, but was observed to fall behind LZ4 Frame in the case with a wide dataset that loaded in many columns (i.e. 19). This is because the wider the dataset is, the more cores it can leverage during the (de)compression phase. However, since deserialization with Bitar can already achieve greater than 100 Gbps throughput that has maxed out the SmartNIC's network bandwidth, the marginal benefit of the additional (de)compression throughput above the NIC's network capability achieved by the resource-intensive software-based approach is minimal considering the limited local storage support on the NIC, especially for tasks focusing on transferring in-transit data. Conservatively speaking, based on these results, the throughput of the compression accelerator rivals that of a software implementation that consumes all the cores of a modern CPU socket. For example, although Bitar's performance is lower than that of LZ4 Frame with 42 threads in the case of testing with the ``Ships'' dataset, it is greater than the same codec's performance with 35 threads when testing with the ``OpenSky Planes'' dataset.

\begin{table}[ht]
\caption{Serialization speedup with Bitar on the host}
\label{tab:serialization_speedup}
\resizebox{\columnwidth}{!}{%
\begin{tabular}{
>{\columncolor[HTML]{EFEFEF}}l lll}
\cline{2-4}
\cellcolor[HTML]{FFFFFF} & \cellcolor[HTML]{FFFFC7}Particles & \cellcolor[HTML]{FFFFC7}OpenSky Planes & \cellcolor[HTML]{FFFFC7}NOAA Ships \\ \hline
LZ4 Frame (single thread)    & 4.61 & 4.71 & 4.71 \\ \hline
Zstd (single thread)         & 7.55 & 7.89 & 8.58 \\ \hline
LZ4 Frame (multiple threads) & 1.39 & 1.44 & 0.95 \\ \hline
Zstd (multiple threads)      & 2.06 & 2.00 & 1.43 \\ \hline
\end{tabular}%
}
\end{table}

\begin{table}[ht]
\caption{Deserialization speedup with Bitar on the host}
\label{tab:deserialization_speedup}
\resizebox{\columnwidth}{!}{%
\begin{tabular}{
>{\columncolor[HTML]{EFEFEF}}l lll}
\cline{2-4}
\cellcolor[HTML]{FFFFFF} & \cellcolor[HTML]{FFFFC7}Particles & \cellcolor[HTML]{FFFFC7}OpenSky Planes & \cellcolor[HTML]{FFFFC7}NOAA Ships \\ \hline
LZ4 Frame (single thread)    & 4.46 & 3.30 & 4.59 \\ \hline
Zstd (single thread)         & 10.84 & 9.51 & 10.20 \\ \hline
LZ4 Frame (multiple threads) & 1.78 & 1.13 & 0.65 \\ \hline
Zstd (multiple threads)      & 2.72 & 2.52 & 1.45 \\ \hline
\end{tabular}%
}
\end{table}

\subsubsection{Impact on Compression Ratio}

Our third question focuses on quantifying how the compression ratio changes when switching between different configurations of the software- and hardware-based compression methods. The compression ratio is computed by dividing the compressed IPC buffer size for a particular configuration by the uncompressed IPC buffer size. We expect the ratio to change in the Bitar hardware implementation because (1) a different compression algorithm is used and (2) the implementation applies compression on the entire table instead of individual columns.

\begin{figure}[ht]
  \includegraphics[width=\columnwidth]{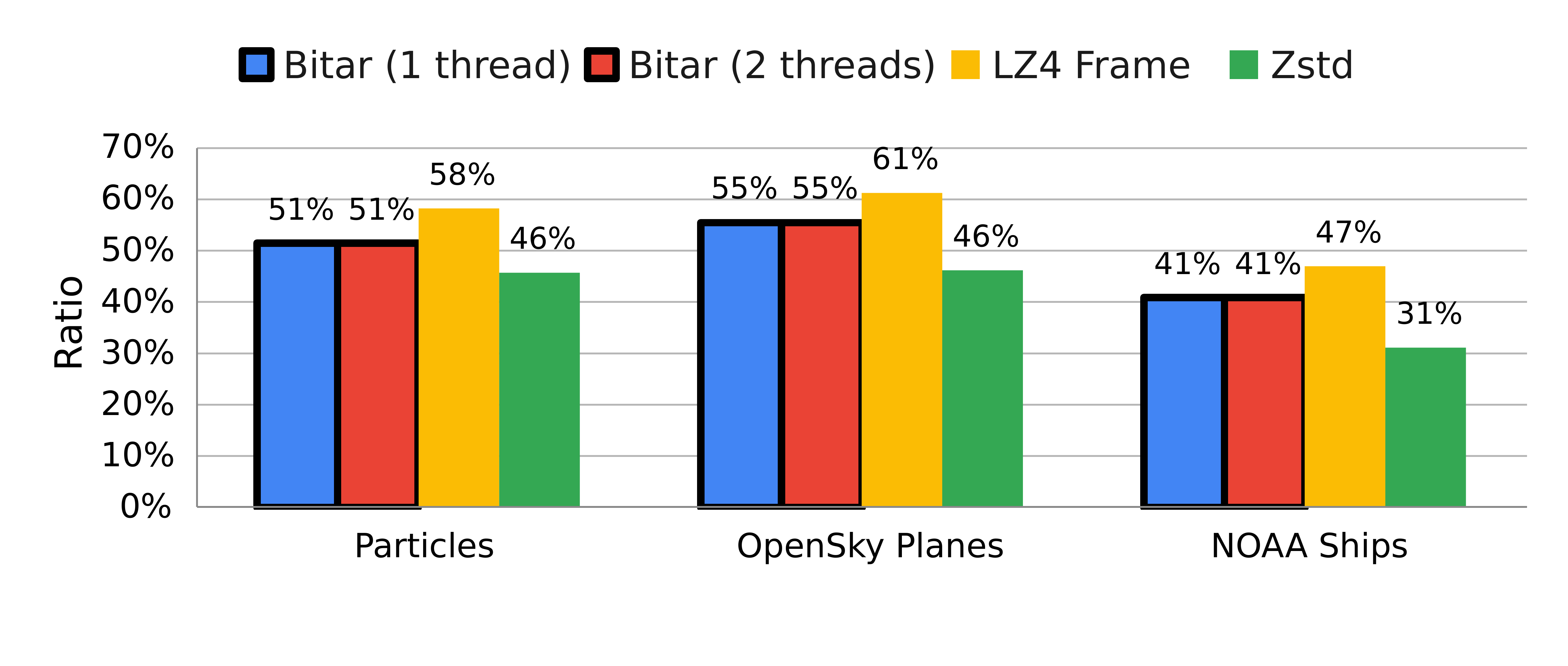}
  \caption{Compression ratios under different compression approaches. Thick black borders indicate hardware (de)compression results.}
  \label{fig:compression_ratios}
\end{figure}

The compression ratios for different configurations are presented in Fig.~\ref{fig:compression_ratios}. Results listed for Bitar are presented for one and two threads to illustrate that splitting the work into multiple threads does not have a significant impact on output size. The hardware-based compression using the DEFLATE algorithm provides a compression ratio that is between that of the LZ4 frame and Zstd codecs in all three datasets. These measurements confirm that offloading computations to the BlueField-2's compression accelerator does not result in a significant sacrifice in the compression ratio.

\subsection{Discussion}

These performance results reveal that general-purpose CPUs are not particularly efficient in (de)compression tasks as the single-thread performance is far lower than that accelerated by compression hardware. Moreover, (de)compression using general-purpose cores cannot effectively scale the performance with the degree of parallelism. In the database arena, recent applications have begun to advocate the use of specialized storage devices that can perform transparent (de)compression to optimize throughput and latency~\cite{cao2020polardb, xiong_computational_2021}. We believe that similar efforts should be made to improve the performance of in-transit data processing. That is, instead of occupying an entire modern CPU socket to gain optimal (de)compression performance, applications can benefit more from running complex logic on these general-purpose cores and offloading compression tasks to hardware accelerators deployed along the data path. For distributed data analytics, having the ability to (de)compress data at near network speeds and with only a fraction of the system's available compute cores is essential for streaming data across nodes.

\section{Summary and Future Work}

While current-generation SmartNICs are slower at processing data than servers, they can perform fundamental data-sifting tasks that are commonly required by different workflows. The compression hardware is particularly appealing for this work, as it provides a way for users to efficiently unpack, process, and repack in-transit data products. However, the current interface for accessing the hardware is challenging to leverage and an obstacle for developers. We present Bitar as a reusable library for simplifying compression on the BlueField-2 cards.

Apache Arrow provides a data model and a collection of operators that are particularly well-suited for processing data on embedded devices that are part of a eusocial processing environment. Arrow's tabular notation allowed us to devise a general framework for storing and processing particle data that did not need to be adjusted when switching between datasets. We note that other types of data may not map to a tabular form as elegantly.

There are multiple paths forward from this work. Having completed the on-card processing work we will transition to network tasks related to distributing data between SmartNICs and coordinating resource utilization across a distributed system. Based on the TCP bottlenecks observed in previous work, it is imperative that these operations take place with RDMA primitives. For the compression work, Arrow will need minor adjustments to allow general users to take advantage of Bitar. These adjustments include modifying Arrow's IPC format to support the DEFLATE codec, incorporating Bitar into Arrow's list of approved third-party libraries, and updating Arrow to route data through Bitar when appropriate. These changes would allow finer-grained access to Arrow data than our current work, as the compression would be applied to individual columns instead of serialized tables.

\section*{Acknowledgment}

This material is based upon work supported by the U.S. Department of Energy, Office of Science, Office of Advanced Scientific Computing Research under Field Work Proposal Number 20-023266. Sandia National Laboratories is a multimission laboratory managed and operated by National Technology \& Engineering Solutions of Sandia, LLC., a wholly owned subsidiary of Honeywell International Inc., for the U.S. Department of Energy's National Nuclear Security Administration under contract DE-NA0003525. This paper describes objective technical results and analysis. Any subjective views or opinions that might be expressed in the paper do not necessarily represent the views of the U.S. Department of Energy or the United States Government.

\bibliographystyle{IEEEtran}
\bibliography{conference_101719}

\end{document}